\documentclass[prd,
preprint,nofootinbib%
 ,secnumarabic%
,amssymb, amsmath,mathrsfs,nobibnotes,aps,11pt]{revtex4}

\usepackage{epsfig}
\usepackage{graphics}
\usepackage{float}
\usepackage[toc,page]{appendix}
\expandafter\ifx\csname package@font\endcsname\relax\else
 \expandafter\expandafter
 \expandafter\usepackage
 \expandafter\expandafter
 \expandafter{\csname package@font\endcsname}%
\fi

\begin{document}
 
%%%%%%%%%%%%%%%%%%%%%%%%%%%%%%%%%%%%%%%%%%%%%%%%%%%%%%%%%%%%%%%%%%%%%%%%%%%%%%%%%%%%%%%%%%%%%%%%%%%%%%%

\title{Constraining Minimal $U(1)_{B-L}$ model from Dark Matter Observations}
\author{Tanushree Basak}
\email{tanu@prl.res.in}
\author{Tanmoy Mondal}
\email{tanmoym@prl.res.in}
\affiliation{Theoretical Physics Division, Physical Research Laboratory, Ahmedabad 380009, India.}

% \emailAdd{tanu@prl.res.in}
% \emailAdd{tanmoym@prl.res.in}
\def\be{\begin{equation}}
\def\ee{\end{equation}}
\def\al{\alpha}
\def\bea{\begin{eqnarray}}
\def\eea{\end{eqnarray}}
%%%%%%%%%%%%%%%%%%%%%%%%%%%%%%%%%%%%%%%%%%%%%%%%%%%%%%%%%%%%%%%%%%%%%%%%%%%%%%%%%%%%%%%%%%%%%%%%%%%%%%%%%%%%%%%

\begin{abstract}
% \abstract{
We study the $B-L$ gauge extension of the Standard Model which contains a singlet scalar and three right-handed neutrinos.
The vacuum expectation value of the singlet scalar breaks the $U(1)_{B-L}$ symmetry. 
Here the third generation right-handed neutrino is qualified as the dark matter candidate, as an artifact of $Z_2$-charge assignment. Relic abundance 
of the dark matter is consistent with WMAP9 and PLANCK data, only near scalar resonances where dark matter mass is almost half of the scalar boson masses.
Requiring correct relic abundance, we restrict the parameter space of the scalar mixing 
angle and mass of the heavy scalar boson of this model. 
Besides this, the maximum value of spin-independent scattering cross-section off nucleon is well-below the {\sc Xenon100} and recent LUX exclusion limits and can be probed by future {\sc Xenon1T} experiment.  
In addition, we compute the annihilation of the dark matter into two photon final state in detail and compare with the Fermi-LAT upper bound on $\langle\sigma v\rangle_{\gamma \gamma}$ for NFW and Einasto profile. 
%}

\end{abstract}
\maketitle

%%%%%%%%%%%%%%%%%%%%%%%%%%%%%%%%%%%%%%%%%%%%%%%%%%%%%%%%%%%%%%%%%%%%%%%%%%%%%%%%%%%%%%%%%%%%%%%%%%%%%%%%%%%%

% \begin{document}
% \maketitle
% \flushbottom
%%%%%%%%%%%%%%%%%%%%%%%%%%%%%%%%%%%%%%%%%%%%%%%%%%%%%%%%%%%%%%%%%%%%%%%%%%%%%%%%%%%%%%%%%%%%%%%%%%%%%%%%%%%%%%%
\section{Introduction}

The existence of missing mass in the galaxies in the form of matter, namely `Dark matter' (DM) was first proposed by Fritz Zwicky in the 1930s. 
According to the recent observations of the anisotropies in the cosmic microwave background by Wilkinson Microwave Anisotropy Probe 
(WMAP9) \cite{Hinshaw:2012aka} -- Universe consists of 71.4\% of dark energy, 4.6\% of luminous matter and 24\% of 
DM. The DM content of the universe has even increased to 26.8\% with the latest PLANCK results \cite{planck}.  
The most convincing evidence for dark matter on galactic scales comes from the observations of 
the galactic rotation curves \cite{rotation_curve} and bullet clusters \cite{bullet}. The presence of dark matter is also supported by the weak 
gravitational lensing of distant galaxies by foreground structure \cite{lensing} and the weak modulation of strong lensing around
individual massive elliptical galaxies \cite{Metcalf:2003sz}. 

Unfortunately, the concept of dark matter does not find an explanation in 
the framework of the Standard Model (SM). Plenty of extensions of  the SM were proposed with a motivation to introduce a suitable DM candidate. 
Among the plethora of candidates, the weakly interacting massive particles (WIMP) are the popular choice (for review see \cite{Jungman:1995df,Bertone:2004pz,
Bergstrom:2009ib}). A simplest extension of the SM with a real or complex gauge singlet scalar field
\cite{McDonald:1993ex,Burgess:2000yq,Davoudiasl:2004be,Guo:2010hq,Bandyopadhyay:2010cc} 
(for latest update see \cite{Cline:2013gha}) has been extensively studied. The scalar turns out to be 
an appropriate DM candidate, which interacts only with the SM Higgs boson. 
Another possibility includes a renormalizable extension of the SM with a gauge singlet Dirac fermion ($\psi$) along with a gauge singlet scalar ($S$)
\cite{Kim:2008pp,Baek:2012uj,Ettefaghi:2013xi}, known as Singlet Fermionic Dark Matter (SFDM) model. In SFDM, 
the singlet scalar interact with the SM Higgs boson whereas 
$\psi$ becomes the viable DM candidate, which interacts to the SM particles via $S$ only.  On the other hand, neutrino mass generation can be 
linked with DM mass through the radiative seesaw mechanism \cite{Kanemura:2011mw,Okada:2012np,Kajiyama:2012xg}, and the Ma-model \cite{Ma-model}. Among other possibilities, the 
 minimal gauge extension of the SM with $U(1)_{B-L}$, and a discrete symmetry ($Z_2$-parity) has been studied by several authors
\cite{Okada:2010wd,Kanemura:2011vm,Kanemura:2011mw,Okada:2012np,Kajiyama:2012xg,Okada:2012sg} in the context of DM.

In this work, we study the minimal $U(1)_{B-L}$ extension of the SM \cite{Khalil,Basso:2010jm,Basso:2011hn}, with an 
additional $Z_2$-symmetry imposed on the model \cite{Okada:2010wd}. Here, only one of the right-handed (RH) 
 neutrinos being odd under $Z_2$-parity, serves as an excellent DM candidate. 
We obtain effectively a Higgs-portal DM which can annihilate into the SM particles (dominantly into $W^+ W^-$ and $ZZ$) and gives correct 
relic abundance \cite{Hinshaw:2012aka,planck} near resonances where DM mass is almost half of the scalar boson masses. Our primary motivation is to restrict the choice of parameter space of this model, based on various recent experimental results of dark matter like relic abundance, limits on 
spin-independent scattering cross-section etc, which has not been considered in earlier studies. We emphasize that the heavy scalar decay width depends strongly on the scalar mixing angle and hence plays a significant role in determining the relic density. Demanding correct relic abundance we constrain the parameter space of the scalar mixing angle and heavy scalar boson mass. We found that the spin-independent elastic scattering cross-section off nucleon is maximum at a particular value of scalar mixing angle and lies below the {\sc Xenon100} \cite{Aprile:2012nq,Lavina:2013zxa} and the latest LUX \cite{lux} exclusion limits. However the future {\sc Xenon1T} \cite{xenon1T} experiment can further restrict the heavy scalar mass. Using the constraints on scalar mixing angle and heavy scalar mass, we have also calculated the annihilation cross-section into two photon final state ($\langle\sigma v\rangle_{\gamma \gamma}$) 
and finally compare with the upper bound on $\langle\sigma v\rangle_{\gamma \gamma}$ by Fermi-LAT \cite{Fermi-data} for different DM profiles. We observe that the resultant $\langle\sigma v\rangle_{\gamma \gamma}$ coincide with the Fermi-LAT data in the region where DM mass is almost half of the light scalar boson mass, otherwise it is well below the Fermi-LAT bound. Apart from DM phenomenology, neutrino mass can be generated in this model via Type-I seesaw mechanism. Here the lightest neutrino remains massless (because of odd-$Z_2$ parity of one of the RH-neutrinos), which is consistent with the observed oscillation data.

The paper is organized as follows: The next section contains a brief description of the model; 
we present an estimation of the relic density in Section~\ref{sec:relic};  
 the direct detection of the DM has been investigated in Section~\ref{sec:SIcross}; a detail calculation for annihilation into  
two photon final state %and an additional gamma-Z line; 
can be found in Section.~\ref{sec:gammaray}; finally we summarize our results and conclude in 
the last section. Appendix~\ref{app:relic_density} shows the estimation of $w(s)$ required for the calculation of relic abundance.
Appendix~\ref{app:loop_functions} contains the loop functions necessary for calculating the 
cross-sections $\langle\sigma v\rangle_{\gamma \gamma}$. 
  A detail calculation of the total decay width of the heavy scalar boson has been shown in appendix~\ref{app:Heavy_higgs_decay_width}.

%%%%%%%%%%%%%%%%%%%%%%%%%%%%%%%%%%%%%%%%%%%%%%%%%%%%%%%%%%%%%%%%%%%%%%%%%%%%%%%%%%%%%%%%%%%%%%%%%%%%%%%%%%%%%%%%
\section{Model}
\label{sec:model}

In this work, we adopt the minimal $U(1)_{B-L}$ extension of the SM \cite{Khalil,Basso:2010jm,Basso:2011hn}. 
Along with the SM particles, this model contains: a SM singlet $S$ with $B-L$ charge +2, three right-handed 
neutrinos $N_R^i (i=1,2,3)$ having $B-L$ charge -1.  As this $U(1)_{B-L}$ symmetry is gauged, an extra gauge boson $Z^\prime$ is associated
as a signature of the extended symmetry. Once the $B-L$ symmetry is broken spontaneously through the vacuum expectation value ($vev$) of $S$,
this $Z^\prime$ becomes massive. Here, we also impose a $Z_2$ discrete symmetry. We assign $Z_2$ charge +1(or even) for all 
the particles except $N_R^3$ \cite{Okada:2010wd}. This ensures the stability of $N_R^3$ which qualified as a viable DM candidate.
The assignment of $B-L$ charge in this model eliminates the triangular $B-L$  gauge anomalies and ensures the gauge invariance of the theory.

Scalar Lagrangian of this model can be written as, 
\begin{equation}\label{eq:new-scalar_L}
\mathcal{L}_s=\left( D^{\mu} \Phi\right) ^{\dagger} D_{\mu}\Phi + \left( D^{\mu} S\right) ^{\dagger} D_{\mu}S - V(\Phi,S ) \, ,
\end{equation}
where the potential term is, 
\be\label{BL-potential}
V(\Phi,S) = 
m^2\Phi^{\dagger}\Phi + \mu ^2\mid S \mid ^2 + \lambda _1 (\Phi^{\dagger}\Phi)^2 +\lambda _2 \mid S\mid ^4 + \lambda _3 \Phi^{\dagger}\Phi\mid S\mid ^2  \, ,
\ee 
with $\Phi$ and $S$ as the Higgs doublet and singlet fields, respectively.
After spontaneous symmetry breaking (SSB) the two scalar fields can be written as,
\begin{equation}\label{eq:SSB}
\Phi =  \left( \begin{array}{c} 0 \\ \frac{v+\phi}{\sqrt{2}} \end{array} \right)\, , 
	\hspace{2cm} S=\frac{v_{_{B-L}}+\phi'}{\sqrt{2}}\, ,
\end{equation} 
with $v$ and $v_{_{B-L}}$ real and positive. Minimization of eq.~(\ref{BL-potential}) gives 
\bea \label{eq:minimization}
m^2 + 2\lambda_1 v^2 + \lambda_3 v v_{_{B-L}}^2 = 0, \nonumber \\
\mu^2 + 4 \lambda_2 v_{_{B-L}}^2 + \lambda_3 v^2 v_{_{B-L}} = 0.
\eea

To compute the scalar masses, we must expand the potential in eq.~(\ref{BL-potential}) around the minima
in eq.~(\ref{eq:SSB}).
Using the minimization conditions, we have the following scalar mass matrix :
\be \label{eq:mass-matrix}
\mathcal{M} = \left( 
		\begin{array}{lr}
                \lambda_1 v^2  & \frac{\lambda_3 v_{_{B-L}} v}{2}\\
                \frac{\lambda_3 v_{_{B-L}} v}{2} & \lambda_2 v_{_\textrm{B-L}}^2
                \end{array}
               \right)
               = \left( 
		\begin{array}{lr}
                \mathcal{M}_{11} & \mathcal{M}_{12}\\
                \mathcal{M}_{21} & \mathcal{M}_{22}\\
                \end{array}
               \right).
\ee
The expressions for the scalar mass eigenvalues ($m_H>m_h$) are:
\be \label{eq:mass-eigenstates}
m_{H,h}^2 = \frac{1}{2}\bigg[\mathcal{M}_{11}+\mathcal{M}_{22} \pm \sqrt{(\mathcal{M}_{11}-\mathcal{M}_{22})^2+4 \mathcal{M}_{12}^2}\bigg].
\ee
The mass eigenstates are linear combinations of $\phi$ and $\phi^{\prime}$, and written as 
\begin{equation}\label{eq:scalari_autostati_massa}
\left( \begin{array}{c} h\\H\end{array}\right)
     = \left( \begin{array}{cc} \cos{\alpha}&-\sin{\alpha}\\ \sin{\alpha}&\cos{\alpha}\end{array}\right)
	\left( \begin{array}{c} \phi\\\phi'\end{array}\right) \, ,
\end{equation}
where, $h$ is the SM-like Higgs boson.
The scalar mixing angle, $\alpha$ can be expressed as:
\be \label{eq:theta}
\tan(2\alpha) = \frac{2 \mathcal{M}_{12}}{\mathcal{M}_{11} - \mathcal{M}_{22}}=\frac{\lambda_3 v_{_{B-L}} v}{\lambda_1 v^2 - \lambda_2 v_{_{B-L}}^2}.
\ee
Now we can calculate the quartic coupling constants by using eqs.~(\ref{eq:mass-eigenstates},\ref{eq:scalari_autostati_massa} and \ref{eq:theta}). 
\begin{eqnarray}\nonumber
\lambda _1 &=& \frac{m_{H}^2}{4v^2}(1-\cos{2\alpha}) + \frac{m_{h}^2}{4v^2}(1+\cos{2\alpha}),\\ \nonumber
\lambda _2 &=& \frac{m_{h}^2}{4v_{_{B-L}}^2}(1-\cos{2\alpha}) + \frac{m_{H}^2}{4v_{_{B-L}}^2}(1+\cos{2\alpha}),\\ \label{inversion}
\lambda _3 &=& \sin{2\alpha} \left( \frac{m_{H}^2-m_{h}^2}{2\,v\,v_{_{B-L}}} \right).
\end{eqnarray}

In the presence of an extra $U(1)_{B-L}$ gauge theory the SM gauge kinetic terms is modified by
\begin{equation}\label{La}
\mathcal{L}_{_{B-L}}^{K.E} = -\frac{1}{4}F^{\prime\mu\nu}F^\prime _{\mu\nu}\, ,
\end{equation}
where,
\begin{eqnarray}\label{new-fs4}
F^\prime_{\mu\nu} &=& \partial _{\mu}B^\prime_{\nu} - \partial _{\nu}B^\prime_{\mu} \, .
\end{eqnarray}
The general covariant derivative in this model reads as
\begin{equation}\label{cov_der_B-L}
D_{\mu}\equiv \partial _{\mu} + ig_S T^{\alpha}G_{\mu}^{\phantom{o}\alpha} 
+ igT^aW_{\mu}^{\phantom{o}a} +ig_1YB_{\mu} +i(\widetilde{g}Y + g_{_{B-L}}Y_{B-L})B'_{\mu}\, .
\end{equation}
Here, we consider only the `pure' $B-L$ model, that is defined
by the condition $\tilde{g} = 0$ at Electro-Weak (EW) scale. This implies zero
mixing at tree level between $Z^\prime$ and Z bosons. 

The relevant Yukawa coupling to generate neutrino masses is given by,
\begin{eqnarray}
\label{yuk}
 \mathcal{L}_{int} &=& \sum_{\beta=1}^3 \sum_{j=1}^2 y_\beta^j \overline{l_\beta}\tilde{\Phi}N_j -  
\sum_{i=1}^3\frac{y_{n_i}}{2}\overline{N_R^i} S N_R^i
\end{eqnarray}
where, $\tilde{\Phi}=-i\tau_2 \Phi^*$.\\
The neutrino mass can be generated in this model via Type-I seesaw mechanism, where the mass matrices for light and heavy 
neutrino are given as,
  \begin{eqnarray}
   m_{\nu_L} &\simeq & m_D^T\; m_M^{-1}\; m_D,\\
   m_{\nu_H} &\simeq & m_M
  \end{eqnarray}
where, $m_D = (y_\beta^j /\sqrt{2})v$ , $(j=1,2)$ and $m_{M_i}=-(y_{n_i}/\sqrt{2} ) v_{_{B-L}}$ , $(i=1,2,3)$.

  Because of $Z_2$-parity, $N_R^3$ has no Yukawa coupling with the left-handed lepton doublet, therefore the lightest neutrino remains massless. 
The masses of $N_R^1$ and $N_R^2$ are considered to be heavier than that of $N_R^3$.

%%%%%%%%%%%%%%%%%%%%%%%%%%%%%%%%%%%%%%%%%%%%%%%%%%%%%%%%%%%%%%%%%%%%%%%%%%%%%%%%%%%%%%%%%%%%%%%%%%%%%%%%%%%%%%%%%%%%

\section{Relic Density}
\label{sec:relic}

In the early universe when the temperature was high enough, the DM particles were in thermal equilibrium with the rest of the cosmic plasma and 
its number density had fallen off exponentially with temperature. But as temperature dropped down below the 
DM mass, the annihilation rate decreased and became smaller than the Hubble expansion rate. Then the DM species was decoupled from the 
cosmic plasma and number density experienced a ``freeze-out'' - hence we observe a significant relic abundance of DM today.  

\begin{table}[h]
\label{tab:parameter}
\begin{center}
\begin{tabular}{|c|c|c|c|} \hline
$m_h$  & $\Gamma_h$  & $v_{_{B-L}}$ &  $g_{_{B-L}}$ \\\hline
      125 GeV     & 4.7$\times 10^{-3}$ GeV  &   7 TeV   &    0.1  \\\hline  
 \end{tabular}
\caption{ Choice of Parameters}
\end{center}
\end{table}

\begin{figure}[t!]
\vspace*{10 mm}
\begin{center}
\includegraphics[scale=0.37,angle=-90]{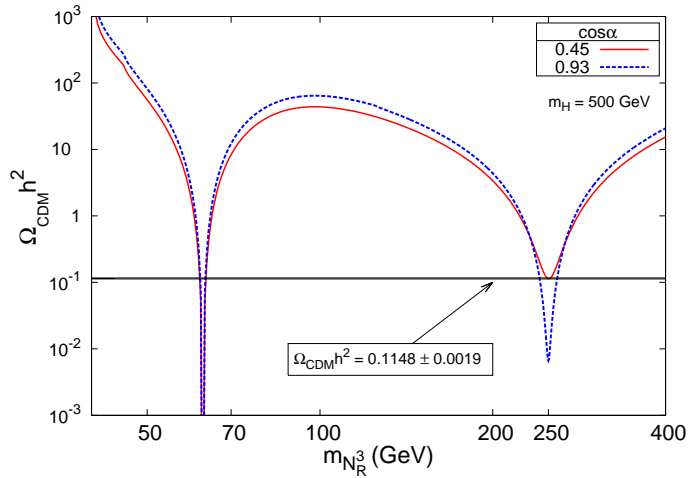}
\vspace*{3mm}
\caption{Plot of relic abundance as a function of DM mass for $m_H=500$ GeV with specific choices of scalar mixing angle $\cos\alpha=0.935 \;
\textrm{(blue-dashed)}, 0.45 \;\textrm{(red-solid)}$. The straight line shows the WMAP9 value, $\Omega_{_{CDM}}h^2=0.1148\pm 0.0019$. }
\label{fig:relic}
\end{center}
\end{figure}

In this model, the right-handed neutrino $N_R^3$ turns out to be a viable dark matter candidate as an artifact of the $Z_2$ charge assignment. We choose 
a specific set of benchmark values for (mass ($m_h$) and decay width ($\Gamma_h$) of SM-like Higgs boson, $vev$ of singlet scalar $S$ and $U(1)_{B-L}$ gauge coupling) our calculation, shown in Table.1, based on present experimental constraints 
 \cite{pdg2012}. However, the mass of the heavy scalar and the scalar mixing angle are not fixed.

The relic abundance of DM can be formulated as \cite{Kolb_Turner},  
\be\label{relic_density}
\Omega_{_{CDM}} h^2 = 1.1\times 10^9 \frac{x_f}{\sqrt{g^*}m_{Pl}\langle\sigma v\rangle_{ann}} \textrm{GeV}^{-1}\; , 
\ee
where $x_f =m_{N_{R}^3}/T_D$ with $T_D$ as decoupling temperature. $m_{Pl}$ is Planck mass  =  $1.22\times 10^{19}$ GeV, and, $g^*$ is effective
number of relativistic degrees of freedom (we use, $g^* = 100 $ and $x_f=(1/20)$). $\langle\sigma v\rangle_{ann}$ is the thermal averaged value of DM annihilation
cross-section times relative velocity. DM interacts with the SM particles via $Z^\prime$-boson and $h,H$. But, $Z^\prime$-boson being heavy ($m_{Z^\prime} \geq 2.33$ TeV \cite{pdg2012}), the 
annihilation of DM into the SM particles takes place via $h$ and $H$ only. Thus, effectively we obtain a Higgs-portal DM model.

$\langle\sigma v\rangle_{ann}$ can be obtained using the well known formula \cite{Srednicki:1988ce}, 
\be\label{thermal_avg}
\langle\sigma v\rangle_{ann} = \frac{1}{m_{N_R^3}^2}\left.\bigg\{ w(s) - 
\frac{3}{2}\Big(2 w(s) - 4{m_{N_R^3}^2} w'(s)\Big) \frac{1}{x_f} \bigg\} \right |_{s=\big(2m_{N_R^3}\big)^2}\; ,
\ee
where prime denotes differentiation with respect to $s$ ($\sqrt{s}$ is the center of mass energy). Here, the function $w(s)$ (detail calculation in appendix~\ref{app:relic_density}) depends on amplitude of different annihilation processes,
\be\label{Dm_decay_channel}
{N_{R}^3} {N_{R}^3} \longrightarrow b\bar{b},\,\tau^+ \tau^-,\, W^+ W^-,\; Z Z,\;\; hh.
\ee

\begin{figure}[t!]
\begin{center}
\includegraphics[scale=0.3,angle=-90]{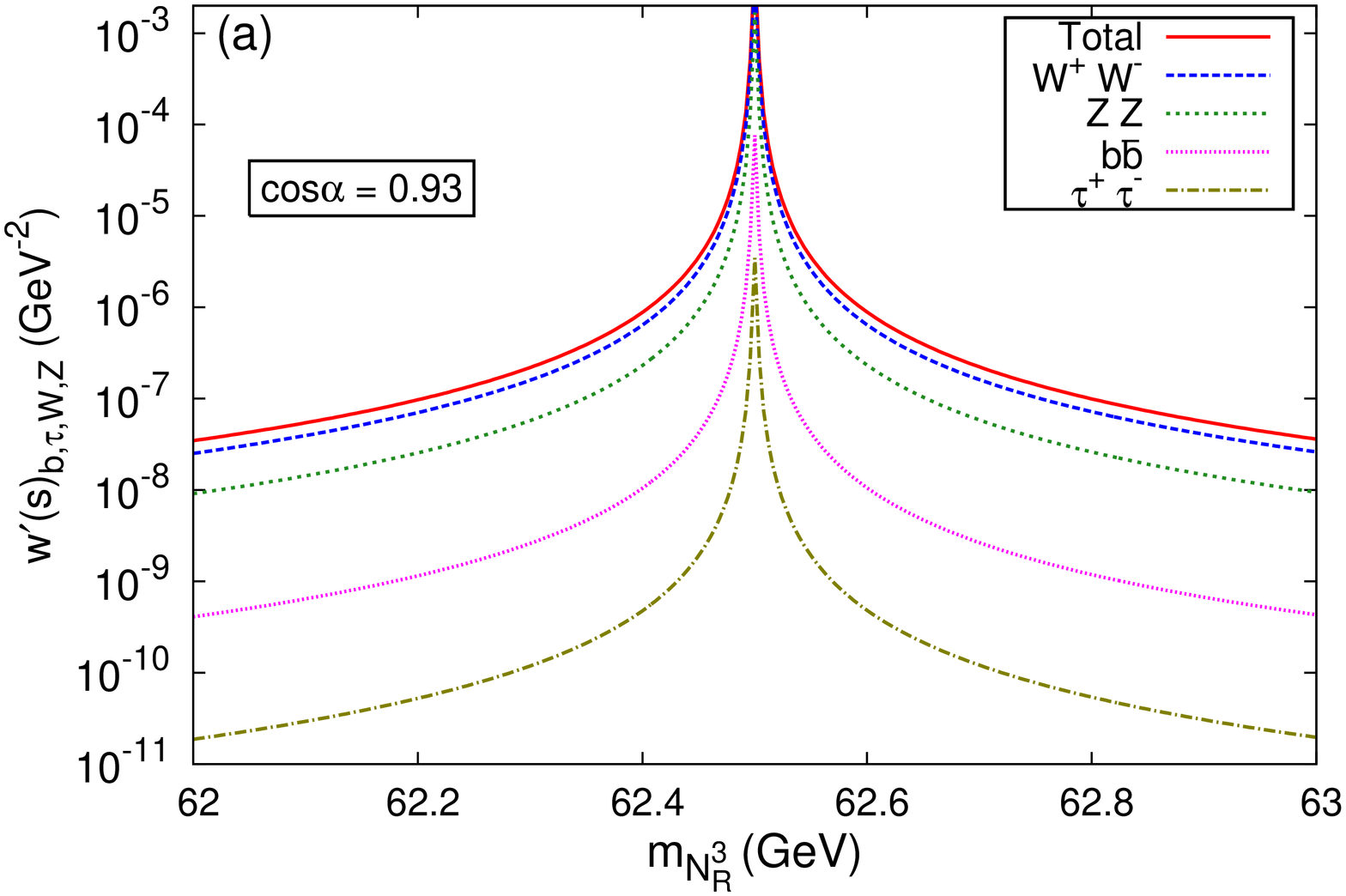}
\includegraphics[scale=0.3,angle=-90]{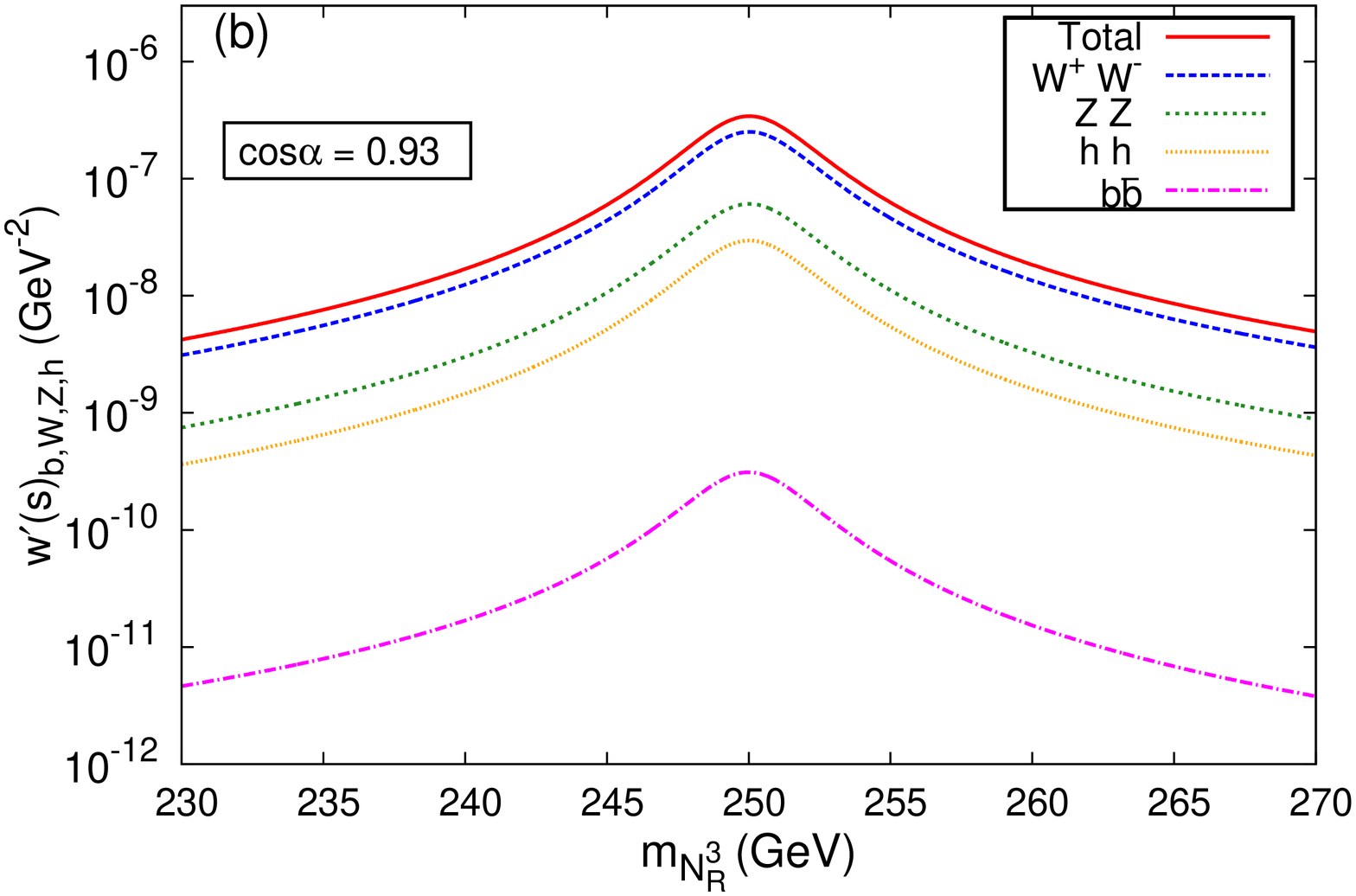}
\caption{ Variation of $w^\prime(s)$ near resonances : (a) $m_{N_{R}^3}=m_{h}/2$  and (b) $m_{N_{R}^3}=m_{H}/2$ , with 
$m_h=125$ GeV and $m_H=500$ GeV, respectively.}
\label{fig:ws}
\end{center}
\end{figure}

In Figure.~\ref{fig:relic} the relic density is plotted against DM mass for two specific choices (to be explained later in this section) of 
scalar mixing angles $\cos\alpha = 0.935$, $0.45$ with $m_H= 500$ GeV. The straight line shows the latest 9-year WMAP data 
i.e, $\Omega_{_{CDM}} h^2 = 0.1148\pm 0.0019$ \cite{Hinshaw:2012aka} 
(whereas latest PLANCK result is, $\Omega_{_{CDM}} h^2=0.1199 \pm 0.0027$ at $68\%$ CL \cite{planck}).
The resultant relic abundance is found to be consistent with the reported value of WMAP-9 and PLANCK experiment 
 only near resonance when, $m_{N_{R}^3} \sim (1/2)\; m_{h,H}$ \footnote{In principle, $Z^\prime$ resonance can also provide the correct relic abundance, but in that case 
the DM mass will be $\mathcal{O}$(TeV) (i.e $m_{N_{R}^3} \sim (1/2)\; m_{Z^\prime}$), if we consider the current experimental bound on $Z^\prime$ mass \cite{pdg2012}.}. 
 The reason for the over abundance of DM except at the resonance can be understood in the following way : The 
annihilation cross-section of DM, being proportional to $y_{n_3}^2$ (where, $y_{n_3}=(\sqrt{2}m_{N_R^3})/v_{_{B-L}}$), 
is heavily suppressed due to large value of $v_{_{B-L}}$.  
Figure.~\ref{fig:relic} also exhibits a strong dependence on the mixing angle near the second resonance (i.e, $m_{N_{R}^3} \sim (1/2)\; m_{H}$). Since, the criterion for correct relic abundance is satisfied near scalar resonances, we have studied the contribution of different annihilation channels to the total annihilation cross-section in that region. We have plotted in Figure.~\ref{fig:ws} the variation of $w^\prime(s)$ ($\langle\sigma v\rangle_{ann}$ depends on $w^\prime(s)$ as shown in eq.~(\ref{thermal_avg})) near resonances $m_{N_{R}^3}=m_{h,H}/2$ for different annihilation channels like $b\bar{b},\,\tau^+ \tau^-,\, W^+ W^-,\, Z Z,\, hh$. 
  We observe that the dominant contribution to the total annihilation cross-section comes from the $W^+ W^-$ , $Z Z$ 
(also final state $hh$ dominance observed in Figure.~2(b)) final states, which is expected because of large SU(2) gauge coupling. In case of Figure.~2(a) a sharp (narrow) 
resonance peak is observed, whereas figure.~2(b) has a broad resonance due to larger decay width ($\Gamma_H$) 
of the heavy scalar, which also depends on scalar mixing angle (see appendix~\ref{app:Heavy_higgs_decay_width}). 

\begin{figure}[t!]
\begin{center}
\vspace*{10 mm}
\includegraphics[scale=0.37,angle=-90]{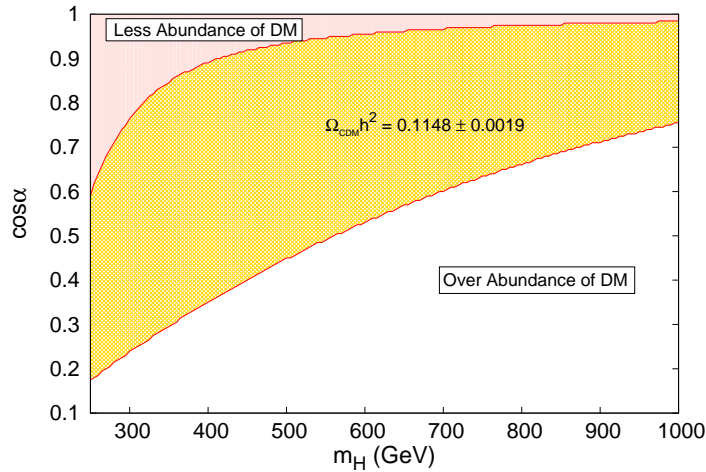}
\vspace*{3mm}
\caption{Yellow region (in the middle) shows the allowed range of $\cos\alpha$ and $m_H$ consistent with correct relic abundance as reported by WMAP9. 
The above-pink (below-white) region is disallowed due to under-abundance (over-abundance) of dark matter.}
\label{fig:relic_scan}
\end{center}
\end{figure}

Relic abundance near the second resonance depends on the following model parameters (unknown) : scalar mixing angle ($\alpha$), heavy scalar mass ($m_H$) and decay width ($\Gamma_H$). But, these are not independent as $\Gamma_H$ can be derived using $\cos\alpha$ and $m_H$. For large mixing angle, the total decay width of heavy scalar is large and hence the  
annihilation cross-section $\langle\sigma v\rangle_{ann}$ is less compared to that with minimal mixing scenario. This behavior is observed 
in Figure.~\ref{fig:relic}, where $\Omega_{_{CDM}} h^2$ is large for smaller value of $\cos\alpha$ (at $m_{N_{R}^3} \sim (1/2)\; m_{H}$) and vice-versa. 
We therefore perform a scan over the entire parameter range of $m_H$ (300-1000 GeV) and $\cos\alpha$ to find the allowed region consistent with 
the 9-year WMAP data ($\Omega_{_{CDM}} h^2 = 0.1148 \pm 0.0019$)\cite{Hinshaw:2012aka}. In Figure.~\ref{fig:relic_scan}, the yellow region shows 
the allowed (by correct relic abundance) range of $\cos\alpha$ for different values of $m_H$, whereas the pink region is forbidden because the 
annihilation cross-section is enhanced for smaller mixing angle (smaller decay width $\Gamma_H$) leading to under-abundance of dark matter. 
On the other hand, the white region is disallowed because of over-abundance. 

% \begin{figure}[t!]
% \begin{center}
% \vspace*{10 mm}
% \includegraphics[scale=0.37,angle=-90]{VS_relic.eps}
% \vspace*{3mm}
% \caption{Yellow region (in the middle) shows the allowed range of $\cos\alpha$ and $m_H$ consistent with correct relic abundance as reported by WMAP9. 
% The above-pink (below-white) region is disallowed due to under-abundance (over-abundance) of dark matter.}
% \label{fig:relic_scan}
% \end{center}
% \end{figure}

%%%%%%%%%%%%%%%%%%%%%%%%%%%%%%%%%%%%%%%%%%%%%%%%%%%%%%%%%%%%%%%%%%%%%%%%%%%%%%%%%%%%%%%%%%%%%%%%%%%%%%%%%%%%%%%%%%%%

\section{Spin-independent scattering cross-section}
\label{sec:SIcross}

% \begin{figure}[t!]
% \begin{center}
% \includegraphics[scale=0.35,angle=-90]{DM_SI_new.eps}
% \caption{ Variation of $\sigma_p^{SI}$ with $m_{N_{R}^3}$ for $m_H=300$ GeV (green-dashed) and $900$ GeV (black-solid) with $\cos\alpha=0.707$. The blue  and violet curves show 
% the bound from {\sc Xenon100} \cite{Aprile:2012nq,Lavina:2013zxa} and LUX \cite{lux} data respectively. Red curve shows the projected limits for {\sc Xenon1T} \cite{xenon1T} .} 
% \label{fig:SICS}
% \end{center}
% \end{figure}

The effective Lagrangian describing the elastic scattering of the DM off a nucleon is given by,
\begin{equation}
 L_{eff}=f_p \bar {{N_{R}^3}}{N_{R}^3} \bar{p}p + f_n \bar {{N_{R}^3}}{N_{R}^3} \bar{n}n \; ,
\end{equation}
where, $f_{p,n}$ is the hadronic matrix element, given by
\begin{equation}
\label{scalarterms}
f_{p,n} = \sum_{q=u,d,s}  f_{Tq}^{(p,n)} a_q  \frac{m_{p,n}}{m_q}  + \frac{2}{27}f_{TG}^{(p,n)}
\sum_{q=c,b,t} a_q \frac{m_{p,n}}{m_q}.
\end{equation}
The f-values are given as in \cite{Ellis:2000ds}
\begin{center}
 $f_{Tu}^{(p)}=0.020 \pm 0.004,\;\; f_{Td}^{(p)}=0.026 \pm 0.005,\;\; f_{Ts}^{(p)}=0.118 \pm 0.062 \; ,$
 \end{center}
 \vspace{-0.5cm}
 \begin{center}
 $f_{Tu}^{(n)}=0.014 \pm 0.003,\;\; f_{Td}^{(n)}=0.036 \pm 0.008$,\;\; $f_{Ts}^{(n)}=0.118 \pm 0.062 \; ,$ 
 \end{center}
and $f_{TG}^{(p,n)}$ is related to these values by
\begin{equation}
f_{TG}^{(p,n)} = 1 - \sum_{q=u,d,s} f_{Tq}^{(p,n)}.
\end{equation}
Here, $a_q$ is the effective coupling constant between the DM and the quark. We obtain 
the scattering cross-section (spin-independent) for the dark matter off a proton or neutron as,
\begin{equation}
\label{eq:sigmaSI}
 \sigma_{p,n}^{SI}=\frac{4m_r^2}{\pi}f_{p,n}^2
\end{equation}
where, $m_r$ is the reduced mass defined as, $1/m_r=1/m_{N_{R}^3}+1/m_{p,n}$.

\begin{figure}[t!]
\begin{center}
\includegraphics[scale=0.35,angle=-90]{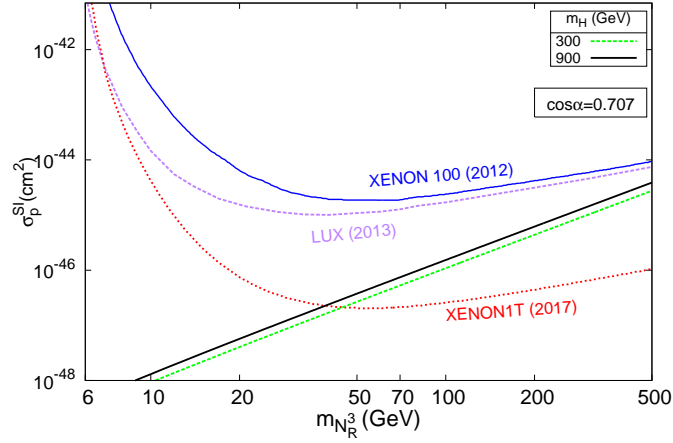}
\caption{ Variation of $\sigma_p^{SI}$ with $m_{N_{R}^3}$ for $m_H=300$ GeV (green-dashed) and $900$ GeV (black-solid) with $\cos\alpha=0.707$. The blue  and violet curves show 
the bound from {\sc Xenon100} \cite{Aprile:2012nq,Lavina:2013zxa} and LUX \cite{lux} data respectively. Red curve shows the projected limits for {\sc Xenon1T} \cite{xenon1T} .} 
\label{fig:SICS}
\end{center}
\end{figure}

An approximate form of $a_q/m_q$ can be recast in the following form :
\begin{eqnarray}
\label{eq:aqmq}
 \frac{a_q}{m_q} &=& \frac{y_{n_3}}{v\sqrt{2}} \Bigg[\frac{1}{m_{h}^2}-\frac{1}{m_{H}^2}\Bigg]\sin\!\alpha \cos\!\alpha \; ,
\end{eqnarray}
%\vspace{-0.3cm}
where, $y_{n_3}=\sqrt{2}m_{N_{R}^3} / v_{_{B-L}}$ is the Yukawa coupling as specified in the second term of eq.~(\ref{yuk}).

From eq.~(\ref{eq:sigmaSI}), it is evident that, $\sigma_{p,n}^{SI}\propto (\sin 2\alpha)^2 f(m_H)$, which is maximum at 
$\alpha=\pi/4$ (or $\cos\alpha=0.707$) irrespective of the choice of $m_H$. Therefore, the maximum value of $\sigma_{p,n}^{SI}$ increases as $m_H$ is increased, which can be understood from eqs.~(\ref{eq:sigmaSI}, \ref{eq:aqmq}). Figure.~\ref{fig:SICS} shows the  
maximum value of spin-independent scattering cross-section (i.e, with $\cos\alpha=0.707$) of the DM off proton ($\sigma_{p}^{SI}$) for $m_H=300$ GeV (green-dashed) and $900$ GeV (black-solid) 
, whereas the blue and violet curves show the {\sc Xenon100} (2012) \cite{Aprile:2012nq,Lavina:2013zxa} and the latest LUX (at 95$\%$ C.L.) 
 \cite{lux} exclusion plots, respectively. The red-curve shows the projected limits on $\sigma_p^{SI}$ for {\sc Xenon1T} experiment \cite{xenon1T}. We observe that the value of the resultant cross-section with two different values of 
 $m_H$ for the entire  range $6\; \textrm{GeV} \leq m_{N_{R}^3} \leq 500 \; \textrm{GeV}$ lies much below the {\sc Xenon100} and latest LUX exclusion limits.
 But, as the value of $m_H$ is increased, the spin-independent cross-section becomes larger at higher values of DM mass and approaches the 
limits as reported by LUX and {\sc Xenon100}. As shown in Figure.~\ref{fig:SICS}, in future {\sc Xenon1T} data might severely restrict the choice of allowed $m_H$.

%%%%%%%%%%%%%%%%%%%%%%%%%%%%%%%%%%%%%%%%%%%%%%%%%%%%%%%%%%%%%%%%%%%%%%%%%%%%%%%%%%%%%%%%%%%%%%%%%%%%%%%%%%%%%%%%%%%%%%%%

\section{Annihilation cross-section into two photons}
\label{sec:gammaray}

The RH-neutrino dark matter $N_{R}^3$ can also annihilate into two photon final state mediated by scalar bosons ($h$ and $H$) 
through loop suppressed processes. Here, we consider mostly dominant contributions from top-quark and W-boson loops to this process 
\cite{Ettefaghi:2013xi}.

The thermal averaging of the annihilation cross-section $\sigma v_{\gamma \gamma}$ can be obtained using \cite{Srednicki:1988ce}
\be\label{eq:gamma_thermal}
\langle\sigma v\rangle_{\gamma \gamma} = \frac{1}{m_{N_R^3}^2}\left.\bigg\{ w(s)_{\gamma \gamma} - 
\frac{3}{2}\Big(2 w(s)_{\gamma \gamma} - 4{m_{N_R^3}^2} w'(s)_{\gamma \gamma}\Big) \frac{1}{x_f} \bigg\} \right |_{s=\big(2m_{N_R^3}\big)^2}\; .
\ee
% where prime denotes differentiation with respect to $s$ ($\sqrt{s}$ is the center of mass energy). 
The function $w(s)_{\gamma \gamma}$ for massless final product is defined as,
\begin{equation}
 w(s)_{\gamma \gamma}=\frac{1}{32\pi} \sum_{spins} |\mathcal{M}_{{N_{R}^3} {N_{R}^3} \rightarrow \gamma \gamma}|^2 .
\end{equation}

Taking into account contributions via $h$ and $H$ bosons we obtain, 
\begin{eqnarray}
 \sum_{spins} |\mathcal{M}_{{N_{R}^3} {N_{R}^3} \rightarrow \gamma \gamma}|^2&=& y_{n_3}^2(s-4m_{N_{R}^3}^2)\bigg \{ 
 \frac{|\mathcal{M}_{h\rightarrow \gamma \gamma}|^2 \sin^2\alpha}{(m_{h}^2-s)^2+m_{h}^2 \Gamma_{h}^2} + 
\frac{|\mathcal{M}_{H\rightarrow \gamma \gamma}|^2 \cos^2\alpha}{(m_{H}^2-s)^2+m_{H}^2 \Gamma_{H}^2} \\ \nonumber && +
\frac{|\mathcal{M}_{h\rightarrow \gamma \gamma}| |\mathcal{M}_{H\rightarrow \gamma \gamma}| \sin\alpha \cos\alpha\{(m_{h}^2-s)(m_{H}^2-s)+m_h m_H \Gamma_h \Gamma_H\}}
{((m_{h}^2-s)^2+m_{h}^2 \Gamma_{h}^2)((m_{H}^2-s)^2+m_{H}^2 \Gamma_{H}^2)}\bigg \} \; .
\end{eqnarray}
where, $\mathcal{M}_{h(H)\rightarrow \gamma \gamma}$ is the amplitude for the decay of $h$($H$) into two photons, which reads as 
\cite{Gunion:1989we,Djouadi:2005gi}
\begin{equation}
 \mathcal{M}_{h(H)\rightarrow \gamma \gamma}= \frac{ g_{_2}\;\alpha_{\textrm{em}}\; m_{h,H}^2 }{8\pi m_W} \Big[3\left(\frac{2}{3}\right)^2 F_t(\tau_t)+F_W(\tau_W)\Big]\cos\alpha(\sin\alpha) \; ,
\end{equation}
where, $\tau_i=4m_i^2/m_{h,H}^2$ ($i=W,\;t$) and $F_{W,\;t}(\tau_{_{W,\;t}})$ are the loop functions for $W$-boson and top-quark respectively 
(see appendix~\ref{app:loop_functions} for detail calculation). $\alpha_{\textrm{em}}$ is the 
electromagnetic fine structure constant at the EW scale,  $\alpha_{\textrm{em}}(m_Z)\sim 1/127$. SU(2) gauge coupling is denoted as $g_{_{2}}$, whereas, $m_W$ is the W-boson mass. 

\begin{figure}[t!]
\label{fermi_data}
 \begin{center}
\includegraphics[scale=0.43,angle=-90]{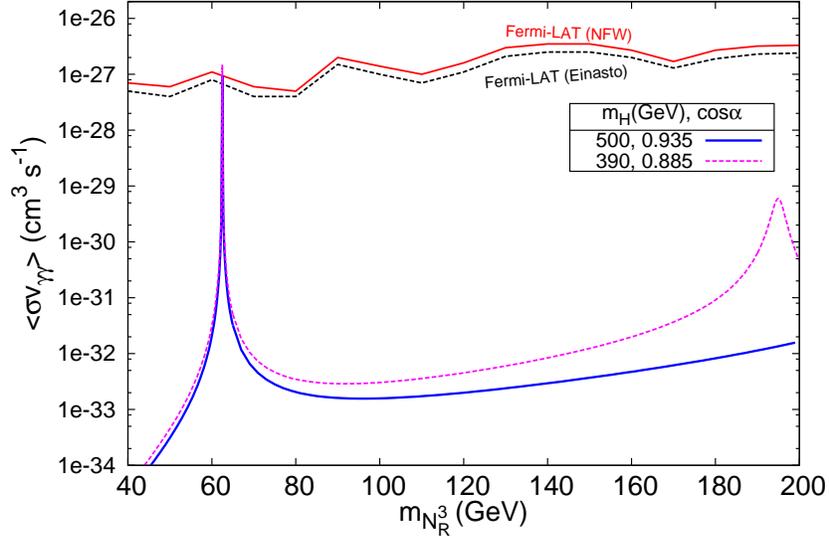}
\caption{Annihilation cross-section into two photon final state vs. dark matter mass with two specific choices : $\cos\alpha=0.935, m_H=500$ GeV (blue-solid) and  $\cos\alpha=0.885, m_H=390$ GeV (purple-dashed) respectively. 
The upper-most two curves show the Fermi-LAT upper bound on  
$\langle\sigma v\rangle_{\gamma \gamma}$ \cite{Fermi-data} for NFW (solid-red) and Einasto (dashed-black) profile. }
\end{center}
\end{figure}  

Figure.~\ref{fermi_data} shows the maximum annihilation cross-section into two photon final state as a function of dark matter mass with different values of $\cos\alpha$ and $m_H$. Here, we have chosen the maximum allowed value of $\cos\alpha$ corresponding to particular value of $m_H$ as derived in Section.~\ref{sec:relic} (see Figure.~\ref{fig:relic_scan}). The blue(pink-dashed) curve shows the resultant $\langle\sigma v\rangle_{\gamma \gamma}$ for $\cos\alpha=0.935 (0.885)$ and $m_H=500 (390)$ GeV. 
It also shows a comparison with the Fermi-LAT upper bound on 
$\langle\sigma v\rangle_{\gamma \gamma}$ for Navarro-Frenk-White (NFW) (solid-red) and Einasto (dashed-black) profile \cite{Fermi-data}. 
We observe a clear coincidence between theoretical plots and Fermi-LAT data near resonance point where $m_{N_{R}^3} \sim (1/2)\; m_{h}$. A second peak is observed in the pink-curve due to a second resonance at $m_{N_{R}^3} \sim (1/2)\; m_{H}$ (i.e. at 195 GeV), but the maximum $\langle\sigma v\rangle_{\gamma \gamma}$ is found to be much below the exclusion limit of Fermi-LAT data. Last year, 
the analysis of Fermi-LAT data \cite{Ackermann:2012qk} had revealed a hint of a monochromatic gamma ray features \cite{Bringmann:2012vr,Weniger:2012tx,Bringmann:2012ez} 
with $E_\gamma\simeq 130$ GeV coming from the 
vicinity of Galactic Center. One of the possible explanations of this phenomena could arise from the annihilation of DM with mass 
$129.8 \pm 2.4^{+7}_{-13}$ GeV and annihilation cross-section $\langle\sigma v\rangle_{\gamma \gamma} = (1.27 \pm 0.32^{+0.18}_{-0.28})
\times 10^{-27} cm^3 sec^{-1}$. It is possible to explain this monochromatic photon line in this model with a 
resonant heavy scalar near 260 GeV and achieve the desired cross-section. But, since the DM dominantly annihilates into $W^+W^-$, $ZZ$ final states ($\langle\sigma v\rangle_{\gamma \gamma}$ is also suppressed as $\mathcal{O}(\alpha_{em}^2(M_Z))$, the continuum photon spectra supersaturate the monochromatic line-like feature.

%%%%%%%%%%%%%%%%%%%%%%%%%%%%%%%%%%%%%%%%%%%%%%%%%%%%%%%%%%%%%%%%%%%%%%%%%%%%%%%%%%%%%%%%%%%%%%%%%%%%%%%%%%%%%%%%%%%%%%%%%%%%%%%%%%%%%%%%%

\section{Summary and Conclusion}
\label{sec:summary}

We have studied a minimal $U(1)_{B-L}$ extended SM, where the third generation RH-neutrino becomes the plausible 
DM candidate by the virtue of an additional $Z_2$-symmetry. The DM considered in this model is effectively Higgs-portal and annihilates dominantly 
into gauge boson ($W^+W^-$, $ZZ$) final states.   
We derive an important constraint on the allowed parameter space of the scalar mixing angle and heavy scalar mass in order
to obtain correct relic abundance. 
Besides this, the relic abundance is found to be consistent with the recent WMAP9 and 
PLANCK data only near scalar resonances, i.e, $m_{N_R^3}=(1/2)\; m_{h,H}$. In future, PLANCK data can further restrict the choice of parameter space. 
The total annihilation cross-section is enhanced due to scalar resonance, 
otherwise it will be suppressed due to heavy $Z^\prime$. The spin-independent elastic scattering cross-section of DM 
off a nucleon is maximum for $\cos\alpha=0.707$, and hence maximum   $\sigma_p^{SI}$ depends on the value of heavy scalar mass. We observe that,  $\sigma_p^{SI}$ is well below the {\sc Xenon100} and LUX exclusion limits for DM mass ranging from $5-500$ GeV. But, future direct detection experiments like {\sc Xenon1T} can put stringent constraint on the choice of $m_H$. The annihilation cross-section of dark matter into $\gamma\gamma$ mediated by 
$h$ and $H$ bosons is compared with that of Fermi-LAT upper bound. We find an agreement between the theoretical plot and the Fermi-LAT data 
near scalar resonance where, $m_{N_R^3}=(1/2)\; m_{h}$.  Although the required $\langle \sigma v_{\gamma \gamma}\rangle$ for explaining 
130 GeV Fermi-line can be obtained in this model, but the gamma-ray continuum spectra produced due to $W^+W^- \;, ZZ$ final state supersaturate this 
monochromatic line feature. In addition, this model can successfully account for the neutrino masses generated via Type-I seesaw mechanism. 
In future, more precise determination of relic abundance and scattering cross-section can be used for obtaining stronger bounds on the allowed 
parameter space of this kind of model.

\section*{Acknowledgements}

We would like to thank Joydeep Chakrabortty, Partha Konar and Subhendra Mohanty for most useful comments and discussions and for their help 
in improving the draft.

\appendices

\section{Calculation of $w(s)$}
\label{app:relic_density}
Let $\phi$ be the scattering angle between incoming DM particles then $w(s)$ can be defined as
\be\label{ws_definition}
w(s) = \frac{1}{32\pi} \sqrt{\frac{s-4 m_{_{final}}^2}{s}}\int \frac{d\!\cos\phi}{2}\sum_{\textrm{all possible channels}}|\mathcal{M}|^2 .
\ee
The function $|\mathcal{M}|^2$ contains not only interaction part, but also contains the kinematical part. 
Considering the processes as in eq.~(\ref{Dm_decay_channel}) we can write
\bea \label{m_squre_sm}
w(s)_{b,\tau,W,Z} &=&\left[ \frac{\sin^2\alpha\cos^2\alpha}{4} \Big(4 y_{n_3}^2(s-4 {m^2_{N_R^3}})\Big)\right]\times \nonumber \\
                & &  \bigg[ \frac{1}{(s-m_h^2)^2+\Gamma_h^2 m_h^2}+ \frac{1}{(s-m_H^2)^2+\Gamma_H^2 m_H^2} \nonumber \\
                & & \hskip 1cm - 2 \frac{(s-m_h^2) (s-m_H^2)+ m_h m_H \Gamma_h \Gamma_H}{{\left((s-m_h^2)^2+
                             \Gamma_h^2 m_h^2\right)\left((s-m_H^2)^2+\Gamma_H^2 m_H^2\right)}}\;\; \bigg] \times \nonumber\\
                & & \bigg[\bigg\{ \frac{1}{8\pi}\sqrt{\frac{s-m_b^2}{s}}   \;4 y_b^2   \; \bigg(\frac{s}{4}-m_b^2\bigg)  3\bigg\} 
                         +\bigg\{ \frac{1}{8\pi}\sqrt{\frac{s-m_\tau^2}{s}}\;4 y_\tau^2\; \bigg(\frac{s}{4}-m_\tau^2\bigg)\bigg\}\nonumber\\
                & & +\bigg\{ \frac{1}{8\pi}\sqrt{\frac{s-m_W^2}{s}}
                   \bigg(\frac{2 m_W^2}{v}\bigg(s+\frac{1}{2m_W^4}\Big(\frac{s}{2}-m_W^2\Big)\bigg)\bigg) \bigg\}\nonumber\\
                & & +\bigg\{ \frac{1}{8\pi}\sqrt{\frac{s-m_Z^2}{s}}
                   \bigg(\frac{ \;\;m_Z^2}{v}\bigg(s+\frac{1}{2m_Z^4}\Big(\frac{s}{2}-m_Z^2\Big)\bigg)\bigg) \bigg\}\bigg] .
\eea
\noindent In this expression second line is the propagator function which includes both $h$ and $H$. Third line shows decay cross section to $b\bar{b}$ and $\tau^+ \tau^-$, 
whereas, fourth and fifth line is decay cross section to $W^+W^-$ ans $ZZ$ respectively. In addition, we have also considered the annihilation 
into the SM-like Higgs bosons, for which $w(s)_h$ is given by,

\bea \label{m_squre_higgs}
 w(s)_{h} &=&\Bigg\{ \frac{1}{16\pi}\Big[4 y_{n_3}^2(s-4 m_{N_R^3}^2)\Big]\sqrt{\frac{s-m_h^2}{s}} \nonumber \\ 
          & &  \bigg(\Big(\frac{\sin\!\alpha}{\sqrt{2}}\Big)^2\frac{\lambda_{hhh}^2}{(s-m_h^2)^2+\Gamma_h^2 m_h^2}  
                    + \Big(\frac{\cos\!\alpha}{\sqrt{2}}\Big)^2  \frac{\lambda_{Hhh}^2}{(s-m_H^2)^2+\Gamma_H^2 m_H^2}  \nonumber \\ 
         & & \hskip 1cm - \frac{\;\sin\!\alpha\;\; \cos\!\alpha\; \;\lambda_{hhh}\;\;\lambda_{Hhh}\;\;\{(s-m_h^2) (s-m_H^2)+ m_h m_H \Gamma_h \Gamma_H\}}
                                {{((s-m_h^2)^2+\Gamma_h^2 m_h^2)\;\;\;((s-m_H^2)^2+\Gamma_H^2 m_H^2)}} \bigg) \Bigg\} \; ,
\eea
where, $\lambda_{hhh}$ and $\lambda_{Hhh}$ are calculated by expanding the Higgs potential part,
\begin{eqnarray}
\lambda_{hhH} &=& 3\lambda_1v\,\left(\cos^2\!\alpha\,\sin\!\alpha\right) + 3\lambda_2\,v_{_\textrm{B-L}}\,\left(\cos\!\alpha\,\sin^2\!\alpha\right) \nonumber \\ 
              & & +\frac{1}{8} \lambda_3\left\{\,v_{_\textrm{B-L}} \left(\cos\!\alpha+3\cos(\!3\alpha)\right)+v \left(\sin\!\alpha-3\sin(\!3\alpha)\right)\right\},\nonumber \\ 
\lambda_{hhh} &=& \frac{\lambda_1}{4}\,v\,\left(3\cos\!\alpha+\cos\!(3\alpha)\right)+\frac{\lambda_2}{4}\,v_{_\textrm{B-L}}\,\left(-3\sin\!\alpha+\sin\!(3\alpha)\right) \nonumber\\
              & & +  \frac{\lambda_3}{8} \left\{v\left(\cos\!\alpha-\cos\!(3\alpha)\right)-v_{_\textrm{B-L}}\left(\sin\!\alpha+\sin\!(3\alpha)\right)\right\} .                 
\end{eqnarray}

Finally, $w(s) = w(s)_{b,\tau,W,Z} + w(s)_{h}$.

\section{Loop functions involved in $\langle\sigma v\rangle_{\gamma\gamma}$}% and $\langle\sigma v\rangle_{\gamma Z}$ }
\label{app:loop_functions}
The loop functions involved in Higgs to di-photon process are depicted as:
\begin{eqnarray}
 F_t(\tau)&=& -2\tau[1+(1-\tau)f(\tau)]\; , \nonumber \\
 F_W(\tau)&=& 2+3\tau +3\tau(2-\tau)f(\tau) \; ,\nonumber
\end{eqnarray}
and
\[f(\tau ) = \left\{ \begin{array}{ll}
{\left( {{{\sin }^{ - 1}}\sqrt {1/\tau } } \right)^2},&\qquad{\rm{for}}\quad\tau  \ge {\rm{1   }}\\
 - \frac{1}{4}\left( {\left. {{\rm{ln}}\frac{{1 + \sqrt {1 - \tau } }}
 {{1 - \sqrt {1 - \tau } }} - i\pi } \right)^2} \right.&\qquad{\rm{for}}\quad\tau {\rm{ < 1 }.}
\end{array} \right.\]

For, $m_h=125$ GeV the loop-functions becomes,
\begin{center}
 $F_t(\tau_t)= 1.83$ , $F_W(\tau_W)=-8.32$.
\end{center}

\section{Calculation for decay width of heavy scalar}
\label{app:Heavy_higgs_decay_width}
In this model we have two Higgs mass eigenstates $(h,H)$ which are admixture of the gauge eigenstates with the mixing angle $\alpha$. The SM gauge eigenstate ($\phi$)
can be written as 
$$
\phi = \cos\!\alpha\;\; h + \sin\!\alpha\;\; H.
$$
So the coupling of $h(H)$ with the SM particles will be multiplied by $\cos\!\alpha(\sin\!\alpha)$. \\

\begin{figure}[t!]
\vspace*{10 mm}
\begin{center}
\includegraphics[scale=0.35,angle=270]{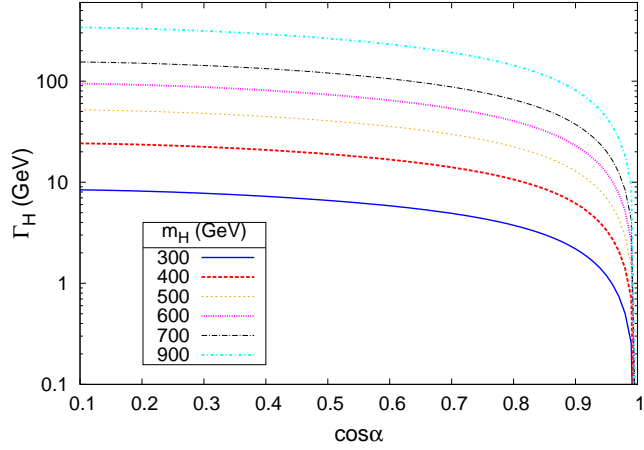}
\vspace*{3mm}
\caption{Plot of heavy scalar boson decay width  as a function of scalar mixing angle $\cos\alpha$} for different values of $m_H$.
\label{fig:gammaH}
\end{center}
\end{figure}

%************************** H----> f fbar ************************************************
\underbar{ Decay of heavy scalar into fermion--antifermion (SM) pair}
\be
\Gamma(H\to f\bar{f}) = N_c \;\frac{\;\;g^2\;m_f^2\;m_H\;\;}{32\;\pi\;m_W^2\;\;}\; \Bigg\{1-\frac{4m_f^2}{m_H^2}\Bigg\}^{3/2}\left(\sin\!\alpha\right)^2
\ee
where $N_c$ is color factor,  1 for leptons and 3 for quarks. 
\\

%************************* H----> W+W- ****************************************************
\underbar{ Decay of heavy scalar into $W$ boson pair}
\be
\Gamma(H\to W^+W^-)=\frac{\;\;g^2\;m_H^3\;\;}{64\;\pi\;m_W^2\;\;}\;\sqrt{1-\frac{4m_W^2}{m_H^2}}\;
\Bigg[1-\frac{4m_W^2}{m_H^2}+\frac{3}{4}\;\left(\frac{4m_W^2}{m_H^2}\right)^2\Bigg]\left(\sin\!\alpha\right)^2
\ee
\\

%************************* H----> ZZ ******************************************************
\underbar{ Decay of heavy scalar into $Z$ boson pair}
\be
\Gamma(H\to ZZ)=\frac{\;\;g^2\;m_H^3\;\;}{128\;\pi\;m_W^2\;\;}\;\sqrt{1-\frac{4m_Z^2}{m_H^2}}\;
\Bigg[1-\frac{4m_Z^2}{m_H^2}+\frac{3}{4}\;\left(\frac{4m_Z^2}{m_H^2}\right)^2\Bigg]\left(\sin\!\alpha\right)^2
\ee

%************************* H----> nu_R nu_R  **********************************************
\underbar{ Decay of heavy scalar into RH neutrinos}
\be
\Gamma(H\to N_R N_R) = \frac{\;m_{N_R}^2\;m_H\;}{16\;\pi\;v_{_\textrm{B-L}}^2}\;\left(1-\frac{4m_{N_R}^2}{m_H^2}\right)^{3/2}\left(\cos\!\alpha\right)^2
\ee

%************************* H----> h h  **********************************************
\underbar{ Decay of heavy scalar into the SM like Higgs}
\be
\Gamma(H\to hh) = \frac{\lambda_{Hhh}^2\;}{32\;\pi\;m_H\;}\sqrt{1-\frac{4m_h^2}{m_H^2}}
\ee

Figure.~\ref{fig:gammaH} shows the dependence of total decay width of the heavy scalar boson $\Gamma_{\textrm{H}}^{tot}$ on the 
scalar mixing $\cos\alpha$ for different values of $m_H$. For higher $m_H$, the decay-width becomes larger for large mixing.
This plot also shows that for the limiting case when $\cos\!\alpha\to1.0$, i.e, without mixing between the scalar bosons, 
$\Gamma_{\textrm{H}}^{tot}\to 0$ and hence it is completely de-coupled from  the SM.

\bibliographystyle{unsrt}
\bibliography{B-L_DM}{}

\end{document}